\documentclass[letterpaper]{article}
\usepackage{aaai25} % DO NOT CHANGE THIS
\usepackage{times} % DO NOT CHANGE THIS
\usepackage{helvet} % DO NOT CHANGE THIS
\usepackage{courier} % DO NOT CHANGE THIS
\usepackage[hyphens]{url} % DO NOT CHANGE THIS
\usepackage{graphicx} % DO NOT CHANGE THIS
\usepackage{graphicx}  % DO NOT CHANGE THIS
\usepackage{natbib}  % DO NOT CHANGE THIS
\usepackage{caption}  % DO NOT CHANGE THIS
\usepackage{amsmath}
\usepackage{amssymb}
\usepackage{subcaption}
\usepackage{placeins}

\newcount\Comments  % 0 suppresses notes to selves in text
\usepackage{color}
\definecolor{darkgreen}{rgb}{0,0.5,0}
\definecolor{purple}{rgb}{1,0,1}
\newcommand{\kibitz}[2]{\ifnum\Comments=1\textcolor{#1}{#2}\fi}
\usepackage{booktabs}

\title{Improving DeFi Accessibility through Efficient Liquidity Provisioning with Deep Reinforcement Learning}

\author{
    Haonan Xu\textsuperscript{\rm 1}, Alessio Brini\textsuperscript{\rm 1}
}

\affiliations {
    \textsuperscript{\rm 1}Pratt School of Engineering, Duke University, Durham, NC, USA\\
    \{haonan.xu@duke.edu, alessio.brini@duke.edu\} 
}

\begin{document}
\maketitle
\begin{abstract}
This paper applies deep reinforcement learning (DRL) to optimize liquidity provisioning in Uniswap v3, a decentralized finance (DeFi) protocol implementing an automated market maker (AMM) model with concentrated liquidity. We model the liquidity provision task as a Markov Decision Process (MDP) and train an active liquidity provider (LP) agent using the Proximal Policy Optimization (PPO) algorithm. The agent dynamically adjusts liquidity positions by using information about price dynamics to balance fee maximization and impermanent loss mitigation. We use a rolling window approach for training and testing, reflecting realistic market conditions and regime shifts. This study compares the data-driven performance of the DRL-based strategy against common heuristics adopted by small retail LP actors that do not systematically modify their liquidity positions. By promoting more efficient liquidity management, this work aims to make DeFi markets more accessible and inclusive for a broader range of participants. Through a data-driven approach to liquidity management, this work seeks to contribute to the ongoing development of more efficient and user-friendly DeFi markets.
\end{abstract}

%%%%%%%%%%%%%%%%%%%%%%%%%%%%%%%%%%%%%%%%%%%%%%%%%%%%%%%%%%%%%%
\section{Introduction}

The emergence of decentralized finance (DeFi)\footnote{DeFi refers to a set of blockchain-based protocols that enable financial activities such as asset exchange, lending, and borrowing without intermediaries, leveraging smart contracts to facilitate transactions transparently and securely.} \cite{zetzsche2020decentralized,auer2024technology} has led to new mechanisms for cryptocurrency exchange that diverge significantly from the traditional limit order book (LOB) models used by centralized exchanges for conventional assets. Within DeFi, decentralized exchanges (DEXs) such as Uniswap v3 implement Automated Market Maker (AMM) mechanisms, where programmable softwares, known as smart contracts, govern both liquidity provisioning and the pricing of token pairs. In contrast to the centralized LOB model, where market makers actively provide liquidity by placing bid and ask orders across a range of prices, DEXs depend on liquidity providers (LPs) who lock assets within liquidity pools \cite{milionis2023complexity}. Trading fees compensate LPs for their commitment, positioning them as central actors in the DeFi ecosystem. While some AMMs, such as dYdX \cite{juliano2018dydx}, incorporate a blockchain-based LOB structure, the majority, including Uniswap, rely on liquidity pools governed by algorithmic pricing mechanisms. Nevertheless, the lack of a traditional LOB in AMMs imposes unique challenges on LPs, who lock their tokens in a given pool, enabling liquidity takers, or swappers, to trade against these assets. This activity shifts prices from the level at the time of liquidity provision, exposing LPs to market volatility and costs that harm profitability.

Uniswap v3 \cite{adams2021uniswap}, an AMM with a constant function market maker (CFMM), introduced the concept of concentrated liquidity \cite{fritsch2021concentrated}, enabling LPs to provide liquidity within specified price ranges rather than across the entire price spectrum. While this feature improves capital efficiency, it also introduces a risk: impermanent loss (IL) can occur not only when the price diverges from its initial value at the inception of the LP position, as observed in earlier versions like Uniswap v2, but also when the market price of the asset pair deviates outside the selected boundaries, rendering the LP’s position inactive and ceasing fee earnings. Concentrated liquidity intensifies losses from price deviations outside the chosen range by removing the offsetting benefit of trading fees. For these reasons, liquidity provision in Uniswap v3 often proves unprofitable, especially for retail traders \cite{loesch2021impermanent}.

A key factor behind this challenge is the frequent arbitrage that occurs between centralized exchanges (CEXs) and decentralized exchanges (DEXs) due to the inherent design of AMMs \cite{wang2022cyclic}. In contrast to the continuous price updates in centralized limit order books, AMMs like Uniswap v3 adjust prices based on the liquidity pool’s token ratio, which can create price staleness when market prices shift rapidly in CEXs and trading is not happening at the same pace on DEXs. This staleness arises because, in the absence of swaps initiated by liquidity takers, the price on blockchain-based AMMs remains static. Since the blockchain lacks knowledge of external market dynamics, it cannot automatically adjust the price according to the price discovery mechanisms occurring off-chain on centralized exchanges. The price staleness exposes LPs to adverse selection costs, particularly through a mechanism called Loss-Versus-Rebalancing (LVR) \cite{milionis2023ammlvr}. LVR, similar to the concept of "picked-off risk" in traditional finance, captures losses incurred when arbitrageurs exploit outdated AMM prices to generate riskless profits by aligning the AMM price with the broader market off-chain. For LPs, especially those using passive, range-bound strategies, this results in realized losses when asset prices deviate from their initial values because they need to modify the price boundaries to ensure they can earn trading fees. The issue is further compounded by the high volatility typical of cryptocurrency markets, which drives LP positions out of profitable ranges, often requiring these costly rebalancing. These factors collectively make liquidity provision a high-risk activity where the fees earned are frequently insufficient to offset losses, posing significant challenges, particularly for smaller and passive LPs that do not modify their liquidity position after deployment.

Managing liquidity provisioning within this AMM framework can be conceptualized as a discrete-time control problem: LPs must decide at each time step whether to retain their current price range at which they deployed their liquidity position on the blockchain or to withdraw and redeploy liquidity at a new range, possibly centered around the current price offered by the AMM. LPs must trade off rebalancing frequency against the potential fees they collect by specifying a narrow price range around the current price. Frequent rebalancing can prevent inactive positions and maintain fee generation, but it incurs gas fee expenses due to withdrawal and redeployment on a blockchain platform and may realize significant losses if the current market price diverges significantly from the LP’s initial position. Additionally, arbitrage opportunities created by price discrepancies between CEXs and DEXs further impact LP profits by contributing to price slippage, amplifying the complexity of liquidity management. 

Optimizing liquidity provisioning within AMMs has implications that extend beyond individual profitability, enhancing the overall value and stability of DeFi systems. Developing more efficient mechanisms for liquidity provision can attract a broader range of participants, including less sophisticated retail traders, by making LP activities more accessible and lowering the barriers posed by high gas fees and frequent arbitrage. DeFi systems foster financial inclusion as they become more active and robust, particularly in regions where traditional banking and financial services remain inefficient or inaccessible \cite{ali2024role}. Ultimately, improving the efficiency of liquidity provision from the LP perspective ensures the sustained growth of the DeFi ecosystem and positions it as a viable and innovative alternative to traditional finance.

The problem of liquidity provisioning naturally unfolds as a sequential decision-making process, where LPs must balance the trade-off between risk and return over time. Reinforcement learning (RL), as a data-driven optimal control framework, is well-suited for addressing such challenges. This paper proposes a deep reinforcement learning (DRL) approach to optimize liquidity provisioning in Uniswap v3, enabling an LP to act as an agent that systematically rebalances its position over time. By modeling LP decision-making through an RL framework, we seek to maximize fees while minimizing losses. The \textit{deep} aspect of DRL leverages neural networks as approximators for value and policy functions, enabling the agent to handle the complexities of liquidity management in dynamic market environments. From this section onward, we refer to RL as the stochastic control problem to be addressed, while DRL denotes the specific algorithmic framework employed to solve it. The active LP agent dynamically adapts its actions based on price movements, balancing the dual objectives of fee maximization and loss mitigation. This approach allows for a data-driven exploration of LPs' optimal rebalancing frequency and positioning strategy, potentially addressing longstanding inefficiencies in passive liquidity provision.

Our contributions are as follows:
\begin{enumerate}
    \item We frame liquidity provisioning in Uniswap v3 as a discrete-time control problem and model LP decision-making using RL, specifically leveraging the Proximal Policy Optimization (PPO) algorithm \cite{schulman2017proximal}. PPO remains one of the best-performing algorithms for tasks involving sequential decision-making. Recent advancements in fine-tuning large language models via RL with human feedback further underscore PPO's prominence in DRL, demonstrating its effectiveness in cutting-edge natural language processing applications \citep{ouyang2022training,zheng2023delve}.
    
    \item We design a reward function that includes fee collection while accounting for the costs associated with deploying liquidity positions and the opportunity cost formalized by LVR. While we incorporate LVR as a penalty to reflect the cost of not investing the same token value in a centralized off-chain exchange, \cite{zhang2023adaptive} does not use LVR in their reward function and instead implements a full rebalancing strategy that would require doubling the token value for liquidity provision. This approach aligns with the objective of developing robust liquidity provision strategies tailored to small retail traders who may lack the capacity to hedge their LP positions continuously.

    \item We propose a state-space design that optimizes both the interval width and its selection as a hyperparameter. By enabling the active LP to dynamically adjust the interval width based on market conditions, our strategy allows for greater flexibility and adaptability in liquidity provision, ensuring effective management of interval positioning and width selection.
\end{enumerate}

The rest of the paper is organized as follows. Sec. \ref{sec:related_works} revisits the related literature, highlighting advancements in decentralized finance and automated market makers. Sec. \ref{sec:uniswap_mechanism} explains in detail the mechanism for providing liquidity in Uniswap v3. Sec. \ref{sec:rl_framework} details the reinforcement learning framework and the PPO algorithm. Sec. \ref{sec:uniswap_environment} discusses the choices made to characterize and implement the Uniswap v3 environment. Sec. \ref{sec:empirical_results} presents and comments on the empirical results. Finally, Sec. \ref{sec:conclusion} concludes the paper.

%%%%%%%%%%%%%%%%%%%%%%%%%%%%%%%%%%%%%%%%%%%%%%%%%%%%%%%%%%%%%%
\section{Related Works}\label{sec:related_works}

This work contributes to two intersecting research domains: DeFi and AMMs, and the integration of machine learning techniques within these frameworks. In this section, we outline the developments within each area, setting the stage for our study’s objectives.

DeFi has rapidly grown over recent years, introducing decentralized, trustless financial systems through smart contracts and blockchain technology \citep{werner2022sok}. Central to this ecosystem are AMM protocols that replace traditional order book models with algorithmically managed liquidity pools. \cite{xu2023sok} provide a comprehensive overview of the mechanics of AMMs. Researchers have expanded on these structures by examining how AMMs incentivize LPs and address risks unique to decentralized markets, such as IL. \cite{aigner2021uniswap,heimbach2022risks,berg2022empirical} explore the risks and inefficiencies associated with LPing in a Uniswap v3 pool. \cite{adams2023costs} takes the opposite perspective and analyzes the costs of swapping on such a platform. \cite{loesch2021impermanent} examines Uniswap v3's leveraged liquidity provision and finds that, across 17 large pools, IL exceeded fee earnings by \$60.8 million, indicating that LPs might have achieved better outcomes by holding their assets. \cite{lehar2021decentralized} analyzes equilibrium in Uniswap’s liquidity pools, showing pool efficiency and limited arbitrage opportunities compared to Binance’s centralized order book.

A specific research direction encompasses studies on optimal liquidity provision strategies, focusing on how LPs can maximize returns while mitigating risks like IL and fluctuating fees in Uniswap-style AMMs. \cite{fan2021strategic} addresses the strategic liquidity provision problem on Uniswap v3, using a neural network framework to optimize earnings by dynamically adjusting liquidity intervals in response to price changes, balancing rewards and gas costs. \cite{wan2022just} examines the "just-in-time" (JIT) liquidity strategy in Uniswap and find that while rare, JIT liquidity enhances trade execution quality and offers price improvements capped at twice the pool's fee rate. Additional quantitative strategies for providing liquidity on Uniswap v3 include those discussed in \cite{fritsch2021concentrated,fan2022differential,bar2023uniswap,cartea2024decentralized}. \cite{zhang2023adaptive} presents a DRL strategy to optimize Uniswap v3 liquidity provision, balancing fee maximization and risk management through adaptive price range adjustments and hedging in centralized futures markets. The paper compares its approach with several of the aforementioned alternative quantitative strategies and incorporates the LVR mechanism as a rebalancing strategy. In contrast, our approach treats LVR as an opportunity cost without implementing its associated rebalancing mechanism. While \cite{zhang2023adaptive} focuses on leveraging hedging through centralized off-chain markets, our strategy exclusively deploys liquidity on-chain, targeting smaller retail traders with limited capacity. By avoiding off-chain hedging, we address the challenges faced by retail LPs, who are known to frequently incur losses when participating in active liquidity provision.

%%%%%%%%%%%%%%%%%%%%%%%%%%%%%%%%%%%%%%%%%%%%%%%%%%%%%%%%%%%%%%

\section{Providing Liquidity in Uniswap v3}\label{sec:uniswap_mechanism}
Each version of Uniswap features independent liquidity pools, with LPs actively choosing to provide liquidity. Specifically, Uniswap v3 operates differently from previous versions of the same protocol, like Uniswap v2, by introducing the concept of \textit{concentrated liquidity}. In this market-making model, LPs specify a price range for providing liquidity. This customization implies that LPs offer their token for swapping only within chosen price ranges instead of spreading the available liquidity over the entire price range. Providing liquidity within narrower intervals where trading is likely increases asset utilization, enabling LPs to earn higher fees relative to their invested capital. Concentrated liquidity improves capital efficiency, but when the price moves outside the range, the LP's position fully converts into the less valuable asset, depending on whether the price exceeds or falls below the range.

In Uniswap v3, LPs choose a price range bounded by specific \textit{ticks}. These ticks, indexed by integers, represent fixed points along the continuous price range which provide the price calculated as $p(i) = 1.0001^i$. Each tick is approximately $0.01\%$ away from the next. Tick indexes are spaced by a predefined interval, termed \textit{tick spacing}, which allows for controlled granularity in price selection. For instance, if the tick spacing is 5, then only multiples of 5 can be chosen as ticks. Different pools on Uniswap have varying tick spacings, which remain fixed and are typically determined by the fee level of the pool \cite{elsts2021liquidity}. Given any price $p_t$ at time $t$, the associated tick index is defined as the largest tick $i$ satisfying $p_t^{'}(i) \leq p_t$. Specifically, the tick index $i$ is computed as:
\begin{equation}\label{eq:tick}
    i = \left\lfloor \frac{\log(p_t)}{\log(1.0001)} \right\rfloor
\end{equation}

Uniswap v3 retains the constant product formula of Uniswap v2, $x_t \cdot y_t = L_{t}^{2}$, where $x_t$ and $y_t$ are the reserves of two tokens $X$ and $Y$ at time $t$, and $L_t$ is the liquidity parameter. However, Uniswap v3 applies the formula dynamically within specified price ranges, rather than across the entire price spectrum as in v2. The liquidity parameter $L_t$ can be interpreted at two levels: at the pool level, it represents the aggregate liquidity provided by all LPs across the pool; at the individual LP level, it reflects the liquidity contribution from a single position. This dual interpretation highlights the modularity of Uniswap v3, where each LP’s contribution can be considered independently while still adhering to the pool's overall constant product constraint. We remark that at the time of deploying a liquidity position, the LP must deposit an amount of tokens $X$ and $Y$ that respect the constant product formula. The spot price $p_t$ of token $X$ in terms of token $Y$ is given by $\frac{y_t}{x_t}$. In terms of the price $p_t$, we can express these reserves as:
\begin{equation}
    x_t = \frac{L_t}{\sqrt{p_{t}}}, \quad y_t = L_t \cdot \sqrt{p_{t}}.
\end{equation}

Uniswap v3 tailors liquidity provision to a specific price range $[p_{t}^{l}, p_{t}^{u}]$, fundamentally altering the constant product formula from Uniswap v2. Within this range, trades adhere to the constant product formula $x_{t}^{v} y_{t}^{v} = L_{t}^{2}$, where $x_{t}^{v}$ and $y_{t}^{v}$ are virtual reserves, which determine trading dynamics without necessarily matching the real reserves of the entire liquidity pool. If the price moves below $p_{t}^{l}$, the LP's position is fully converted to token $X$, and when the price rises above $p_{t}^{u}$, it is fully converted to token $Y$. In both cases, the LP retains the less valuable asset, ceasing trading activity and amplifying impermanent loss (IL).

To accommodate these dynamics, the constant product curve is translated, creating a curve of "real balances" that reflects the actual token holdings of the LP at any price $p_t$. For $p_t \in [p_{t}^{l}, p_{t}^{u}]$, the real balances are calculated as:
\begin{equation}\label{Eq:lpreserves}
    x_t = L_t \Big(\frac{1}{\sqrt{p_t}} - \frac{1}{\sqrt{p_{t}^{u}}}\Big), \quad y_t = L_t (\sqrt{p_t} - \sqrt{p_{t}^{l}}).
\end{equation}
These adjustments ensure that at $p_t = p_{t}^{l}$, the LP holds only token $X$ ($y_t = 0$), and at $p_t = p_{t}^{u}$, only token $Y$ ($x_t = 0$). Outside the range $[p_{t}^{l}, p_{t}^{u}]$, the LP's token balances remain fixed at these respective endpoints, as no trades occur.

In Uniswap v3, IL occurs when the market price $p_t$ deviates from the initial entry price $p_0$ at $t=0$, whether within or outside the specified price range. This risk is magnified in v3 compared to v2 due to the concentrated liquidity model. The IL is quantified as the percentage difference between the value of the LP’s position, $V_{t}(p_{t})$, and the hypothetical value if the tokens had not been deposited, $W_{t}(p_{t})$:
\begin{equation}
    \text{IL}_{t}(p_{t}) = \frac{V_{t}(p_{t})}{W_{t}(p_{t})} - 1,
\end{equation}
where $V_{t}(p_{t}) = x_{t} \cdot p_{t} + y_{t}$ represents the value of the LP's position at price $p_{t}$, and $W_{t}(p_{0}) = x_0 \cdot p_0 + y_0$ is the hypothetical value of holding the initial token amounts $x_0$ and $y_0$ outside the AMM. For further details on the derivation and analysis of IL in Uniswap v3 and the relationship between virtual and real reserves, see \cite{ottina2023automated}.

We calculate trading fees following \cite{fan2021strategic,zhang2023adaptive}, which provide a detailed framework for fee distribution based on price movements. For a price change within the range $[p_{t}^{l}, p_{t}^{u}]$, where $p_{t} \leq p_{t+1}$, the trading fee earned by an LP holding $L_{t}$ liquidity units is given by:
\begin{equation}\label{Eq:fees_up}
    f_{t} = \frac{\delta}{1 - \delta} L_{t} \Big(\sqrt{p_{t+1}} - \sqrt{p_{t}}\Big),
\end{equation}
where $\delta$ represents the fixed fee rate. In the case of downward price movements, where $p_{t} > p_{t+1}$, the fee is computed as:
\begin{equation}\label{Eq:fees_down}
    f_{t} = \frac{\delta}{1 - \delta} L_{t} \Big(\frac{1}{\sqrt{p_{t+1}}} - \frac{1}{\sqrt{p_{t}}}\Big) p_{t+1}.
\end{equation}
If the price movement extends beyond the bounds of the LP's range $[p_{t}^{l}, p_{t}^{u}]$, the total movement can be decomposed into segments, and LPs accrue fees only for price movements within their active range.

%%%%%%%%%%%%%%%%%%%%%%%%%%%%%%%%%%%%%%%%%%%%%%%%%%%%%%%%%%%%%%

\section{Deep Reinforcement Learning with PPO}\label{sec:rl_framework}

In this section, we provide a detailed overview of the RL mathematical framework, establishing its relevance to the active LP agent we develop and train in Sec. \ref{sec:empirical_results}. 

RL operates within a Markov Decision Process (MDP), defined by states $S_t\in\mathcal{S}$, actions $A_t\in\mathcal{A}$, and transition probabilities $\mathcal{P}{ss'}^{a}=P[S{t+1} = s' \mid S_{t} = s, A_{t} = a]$. An RL agent aims to maximize the expected cumulative (discounted) rewards by identifying the optimal action based on the current state, i.e. 
%Hence, it represents the stochastic control problem 
\begin{equation}\label{Eq:rlproblem}
\max_{{\pi}} \mathbb{E}\left[ \sum_{t=0}^\infty \gamma^t R_{t+1}(S_t,A_t,S_{t+1} ) \right] \,,
\end{equation}
where $t$ refers to the timestep at which the agent is making a decision, $\pi$ denotes the agent's strategy, associating a probability $\pi(a \mid s)$ with the action $A_{t}=a$ given the state $S_{t}=s$, and $R_{t+1}$ is the scalar reward that the agent obtains from making an action at time $t$.

The model-free reinforcement learning scenario is particularly suited for the active LP agent, as it reflects the lack of prior knowledge regarding key aspects of the environment. The LP agent lacks prior knowledge of the token pair price dynamics in the liquidity pool where she provides liquidity. Additionally, the agent is unaware of the exact functional form of the reward function, which encapsulates both the costs (e.g., gas fees and impermanent loss) and the benefits (e.g., trading fees). For instance, the active LP agent trained as a model-free RL agent must infer the impact of impermanent loss (IL) on their position and adapt their strategy accordingly, without any predefined model or prior information. Hence, this model-free framework implies that the agent lacks access to the transition probability $\mathcal{P}_{ss'}^{a} = P[S_{t+1} = s' \mid S_t = s, A_t = a]$ governing the environment's dynamics. Instead, the agent learns these dynamics by observing sequences of states, actions, and rewards through interactions with the environment.

RL methods can be broadly categorized into value-based and policy-based approaches. The family of value-based methods is characterized by the action-value function
\begin{equation}\label{Eq:Qfunc}
Q_{\pi}(s, a) \equiv \mathbb{E}\left[\sum_{k=0}^\infty \gamma^k R_{t+1+k} \mid S_{t}=s, A_{t}=a, \pi\right],
\end{equation}
represents the long-term reward associated with taking action $a$ in state $s$ when following the strategy $\pi$ thereafter. Estimating Eq. \eqref{Eq:Qfunc} enables deriving an optimal deterministic policy as the action with the highest value in each state. Depending on how the agent estimates the action-value function, different value-based algorithms can be introduced.

Policy gradient algorithms, such as PPO, offer an alternative approach to solving the optimization problem in Eq. \eqref{Eq:rlproblem}. In this case, the optimal strategy is directly parameterized within a specific policy class $\pi_{\theta} = \pi(A_{t}\mid S_{t}; \theta)$, which can be represented by a multilayer neural network with parameters $\theta$.

To maximize the expected cumulative (discounted) rewards, a policy gradient algorithm computes the gradient of the performance measure \\ $J(\theta) = \sum_{t=0}^\infty \gamma^t R_{t+1}(S_t,A_t,S_{t+1};\pi_{\theta})$, and updates to the parameters of the policy using gradient ascent:
\begin{equation}\label{Eq:PGupdate}
\theta_{t+1}=\theta_t +\alpha \nabla_{\theta} J(\theta_{t}),
\end{equation}
where $\alpha$ represents the learning rate. Unlike value-based methods, which rely on value functions, policy gradient algorithms enable agents to select actions without consulting a value function.

The policy gradient theorem \cite{sutton2000,marbach2001} provides an analytical expression for the gradient of $J(\theta)$
\begin{align}\label{Eq:pgtheorem}
\nabla_{\theta} J(\theta) &= \mathbb{E}_{\pi_{\theta}} \left[ \frac{\nabla_{\theta} \pi\left(A_{t} \mid S_{t};\theta \right)}{\pi\left(A_{t} \mid S_{t};\theta \right)} Q_{\pi_\theta}(S_{t}, A_{t}) \right] \ \nonumber \\
&= \mathbb{E}_{\pi_{\theta}} \left[ \nabla_{\theta} \log\pi\left(A_{t} \mid S_{t};\theta \right) Q_{\pi_\theta}(S_{t}, A_{t}) \right],
\end{align}
where the expectation, with respect to $(S_t,A_t)$, is taken along a trajectory (episode) that follows the policy $\pi_{\theta}$.

Subtracting a baseline $V_{\pi}(s)$ from the action value function $Q_\pi(s,a)$ in Eq. \eqref{Eq:pgtheorem}reduces the variance of the empirical average along the episode without adding bias. A commonly used baseline is the state-value function
\begin{equation}\label{Eq:Vfunc}
V_{\pi}(s) \equiv \mathbb{E}\left[\sum_{k=0}^\infty \gamma^k R_{t+1+k} \mid S_{t}=s, \pi\right],
\end{equation}
which represents the long-term reward starting from state $s$ if the policy $\pi$ is followed afterward. Consequently, the gradient can be rewritten as
\begin{equation}\label{Eq:pgadvantage}
\nabla_{\theta} J(\theta) = \mathbb{E}_{\pi\theta} \left[ \nabla_{\theta} \log\pi\left(A_{t} \mid S_{t};\theta_{t} \right) \mathbb{A}_{\pi\theta}(S{t}, A_{t}) \right],
\end{equation}
where
\begin{equation}\label{Eq:Afunc}
\mathbb{A}_{\pi}(s, a) \equiv Q_{\pi}(s, a)-V_{\pi}(s),
\end{equation}
denotes the advantage function, which quantifies the gain obtained by selecting a specific action in a given state relative to its average value for policy $\pi$.

Policy gradient algorithms differ in how they estimate the advantage function.
In the case of PPO, the advantage estimator $\mathbb{A}\left(s,a;\psi\right)$ is parametrized by another neural network with parameters $\psi$. This approach, known as actor-critic, involves the policy estimator $\pi(a \mid s;\theta)$ (actor), which outputs the mean and standard deviation of a Gaussian distribution used for action sampling, and the advantage function estimator $\mathbb{A}\left(s,a;\psi\right)$ (critic), which provides a scalar value. During the learning process, the two neural networks interact: the critic drives updates to the actor, which then collects new sample sequences to update the critic and evaluate further updates. The PPO algorithm optimizes the extended objective function given by
\begin{equation}\label{Eq:ppoobj}
J^{\text{PPO}}(\theta,\psi)= J(\theta)-c_{1} L^{\text{AF}}(\psi) +c_{2} H\left(\pi\left(a \mid s; \theta\right)\right),
\end{equation}
where the second term represents a loss between the advantage function estimator $\mathbb{A}\left(s,a;\psi\right)$ and a target $\mathbb{A}^{\text{targ}}$ (the cumulative sum of discounted rewards) used to train the critic neural network. The last term introduces an entropy bonus to ensure adequate exploration. 

%%%%%%%%%%%%%%%%%%%%%%%%%%%%%%%%%%%%%%%%%%%%%%%%%%%%%%%%%%%%%%%%%%%%%%%%%%%%%%%%%%%%%
\section{AMM Environment for Active LPs}\label{sec:uniswap_environment}

In this section, we detail the choices for constructing the RL environment to train our active LP agent. This includes state representation, action space design, reward function, and implementation details for creating and training the RL model.

We implement a custom RL environment using the \texttt{Gymnasium}\footnote{\url{https://gymnasium.farama.org/}} package in Python to simulate interactions, and we employ \texttt{Stable Baselines3}\footnote{\url{https://stable-baselines3.readthedocs.io/en/master/}} to train the PPO policy. We use \texttt{Optuna}\footnote{\url{https://optuna.org/}} for hyperparameter optimization. 

The agent's state space integrates immediate and historical market data to capture token pair price dynamics. The state representation includes:
\begin{itemize}
    \item Market price $p_t$.
    \item Tick index $i$ associated with $p_t(i)$ according to Eq. \eqref{eq:tick}.
    \item Interval width $w_t$, defined as a multiple of tick spacing, representing the range of liquidity provision selected in the previous discrete timestep.
    \item Current liquidity level $L_{t}$ of LP assets calculated using Eq. \eqref{Eq:lpreserves}.
    \item Exponentially weighted volatility of the market price $\sigma_t(p_{t})$ calculated with a smoothing factor $\alpha=0.05$.
    \item 24-window and 168-window moving averages to capture recent price movements.
    \item Technical indicators\footnote{We use \texttt{TA-Lib} to compute the technical indicators}, including Bollinger Bands (BB), Average Directional Movement Index Rating (ADXR), Balance of Power (BOP), and Directional Movement Index (DX), provide trend-following and mean reversion signals.
\end{itemize}

The action space, denoted as $\mathcal{A}$, consists of a discrete set of actions, including a 0 action and 2 to 5 different tick size options for the liquidity price interval. These actions determine the width of the symmetric price interval centered around the current market price, which is used for deploying liquidity. For instance, a possible action space could be $\mathcal{A} = \{0, 10, 20, 30, 40\}$. We optimize the action space as a hyperparameter to identify the best choices for the active LP. This setup allows us to explore how different interval widths impact the performance of the RL agent. The optimized action space values for each out-of-sample window are detailed in App. \ref{App}.

The reward function incorporates an LVR penalty adapted from \cite{milionis2023ammlvr}. For a price $p_{t}$ within the liquidity range $[p_{t}^{l}, p_{t}^{u}]$, the value of the pool reserves is calculated according to Eq. \eqref{Eq:lpreserves}. The pool value and the second derivative become
\begin{equation}\label{Eq:lvrvp}
V_{t}(p_{t}) = L_{t} \left(2\sqrt{p_{t}} - \frac{p_{t}}{\sqrt{p_{t}^{u}}} - \sqrt{p_{t}^{l}}\right), \; V_{t}^{''}(p_{t}) = -\frac{L_{t}}{2p_{t}^{3/2}}.
\end{equation}
\cite{milionis2023ammlvr} defines instantaneous LVR as
\begin{equation}\label{Eq:lvrell}
\ell_{t}(\sigma, p_{t}) = \frac{L\sigma^2}{4}\sqrt{p_{t}},
\end{equation}
where we replace the fixed volatility $\sigma$ with the exponentially weighted volatility $\sigma_{t}(p_{t})$.

The reward function includes a penalty for gas fees incurred when deploying liquidity positions. Gas fees apply only when the active LP modifies the range-bounded position; if the position remains unchanged at subsequent timesteps, no gas fees are applied. However, if the active LP chooses a new price range, gas fees are incurred twice: once for withdrawing the liquidity and once for redeploying it. Token-swapping activity while the liquidity position is active generates trading fees, the positive component of the reward function, as described in Eqns. \eqref{Eq:fees_up} and \eqref{Eq:fees_down}. The reward function becomes
\begin{equation}\label{eq:reward}
    R_{t+1} = f_t - \ell_{t}(\sigma, p) - \mathbb{I}[a_t \neq 0] \cdot 
    \begin{cases} 
        g, & \text{if } t = 0 \\ 
        g, & \text{if } t > 0,
    \end{cases}
\end{equation}
where $g$ represents the fixed costs for the gas fees which we set at \$5 based on Etherscan's gas tracker\footnote{\url{https://etherscan.io/gastracker}}.

In initializing each liquidity position, the risky token $X$ quantity is set to $x_{0} = 2$, with the numeraire token $Y$ computed based on the current liquidity level and market price. At each discrete hourly step, the agent decides whether to withdraw and redeploy liquidity with adjusted price boundaries or to maintain the current interval, balancing risk and profitability. The agent may opt to keep the interval unchanged even if the market price moves outside its boundaries, temporarily halting fee earnings. This decision reflects a long-term optimization strategy to manage the riskiness of the LP position while maximizing profitability.
%%%%%%%%%%%%%%%%%%%%%%%%%%%%%%%%%%%%%%%%%%%%%%%%%%%%%%%%%%%%%%
\section{Empirical Studies}\label{sec:empirical_results}

The dataset for this study consists of hourly data spanning from 5 May 2021 at 1:00 am, marking the inception of Uniswap v3, to 29 January 2024 at 7:00 pm. We focus on the WETH/USDC Uniswap pool at the 0.05\% fee tier\footnote{Contract address on Ethereum Blockchain: \url{0x88e6a0c2ddd26feeb64f039a2c41296fcb3f5640}}. We collect data via the Uniswap Ethereum subgraph and resample the unevenly spaced transaction data into an evenly spaced hourly series.

To train the model, we adopt a rolling window approach. Each rolling window includes a training period of 7,500 hours (approximately 10 months) followed by a testing period of 1,500 hours (approximately 2 months). After each iteration, the window is shifted by 1,500 timesteps, and the process is repeated. We optimize the model's hyperparameters during each training window to ensure the best configuration for that dataset period.

\begin{figure}[h!]
    \centering
    \includegraphics[width=0.5\textwidth]{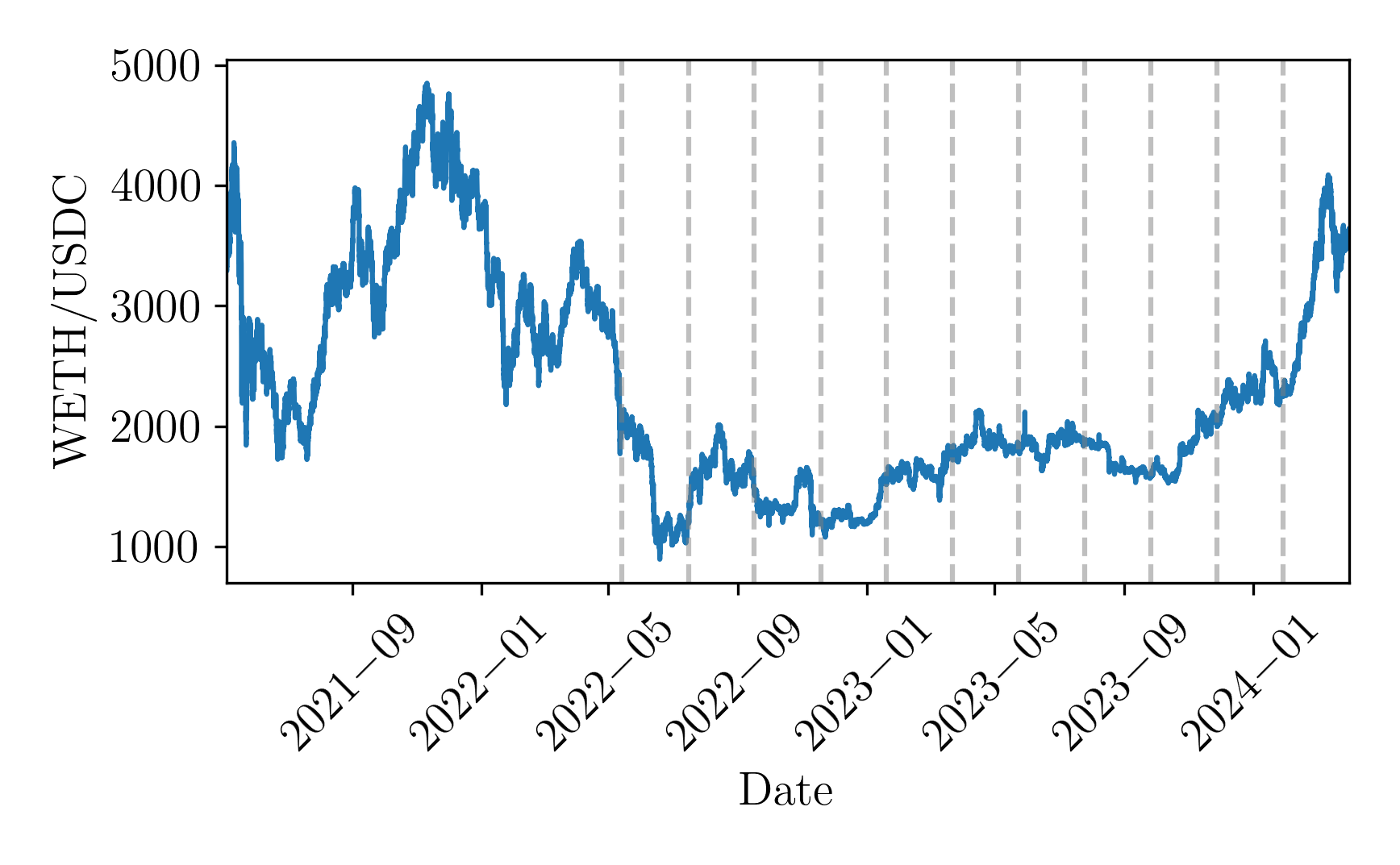}
    \caption{Price series of WETH/USDC and corresponding sliding windows: The figure illustrates the hourly price series of the WETH/USDC token pair in the Uniswap pool with a 0.05\% fee tier, highlighting the sliding windows used for training and testing the RL model.}
    \label{Fig:ethprice}
\end{figure}

Fig. \ref{Fig:ethprice} shows the resampled hourly WETH/USDC price dynamics, with vertical dashed gray lines marking the boundaries of the out-of-sample testing windows. These divisions highlight the distinct patterns captured in different training and testing periods.

We choose this rolling window approach to reflect the practical challenges of deploying DRL models in real-world financial markets, particularly in the volatile cryptocurrency market. Unlike a naive approach that trains over the entire dataset and tests with a single fixed cutoff, our method accounts for potential regime shifts in the market, such as upward or downward trends. Training an RL-based active LP agent on a single window risks overfitting to one market regime, reducing its effectiveness in varying conditions. By contrast, our rolling training method mimics real-world application scenarios, ensuring the model’s adaptability across diverse market patterns. This approach also enables a more realistic comparison with passive LP strategies commonly employed in practice.

%%%%%%%%%%%%%%%%%%%%%%%%%%%%%%%%%%%%%%%%%%%%%%%%%%%%%%%%%%%%%%
%%%%%%%%%%%%%%%%%%%%%%%%%%%%%%%%%%%%%%%%%%%%%%%%%%%%%%%%%%%%%%

For each out-of-sample window, we train 50 different active LP agents, selecting the one that performs best according to the cumulative sum of the reward function described in Eq. \eqref{eq:reward}. We compare the optimized active LP's performance against the passive LP, a periodic strategy that modifies liquidity positions at fixed 500-hour intervals (approximately 20 days), to evaluate adaptability to market dynamics. This periodic adjustment is a heuristic approach that recenters the liquidity range at evenly spaced intervals, without adapting to market dynamics.

We set the tick width for the passive LP's price interval to 50 ticks above and below the current price. In contrast, the active LP dynamically adjusted its price interval and rebalancing frequency based on learned policies optimized through reinforcement learning.

\begin{figure}[h!]
    \centering
    \begin{subfigure}[b]{0.48\linewidth}
        \centering
        \includegraphics[width=1.1\linewidth]{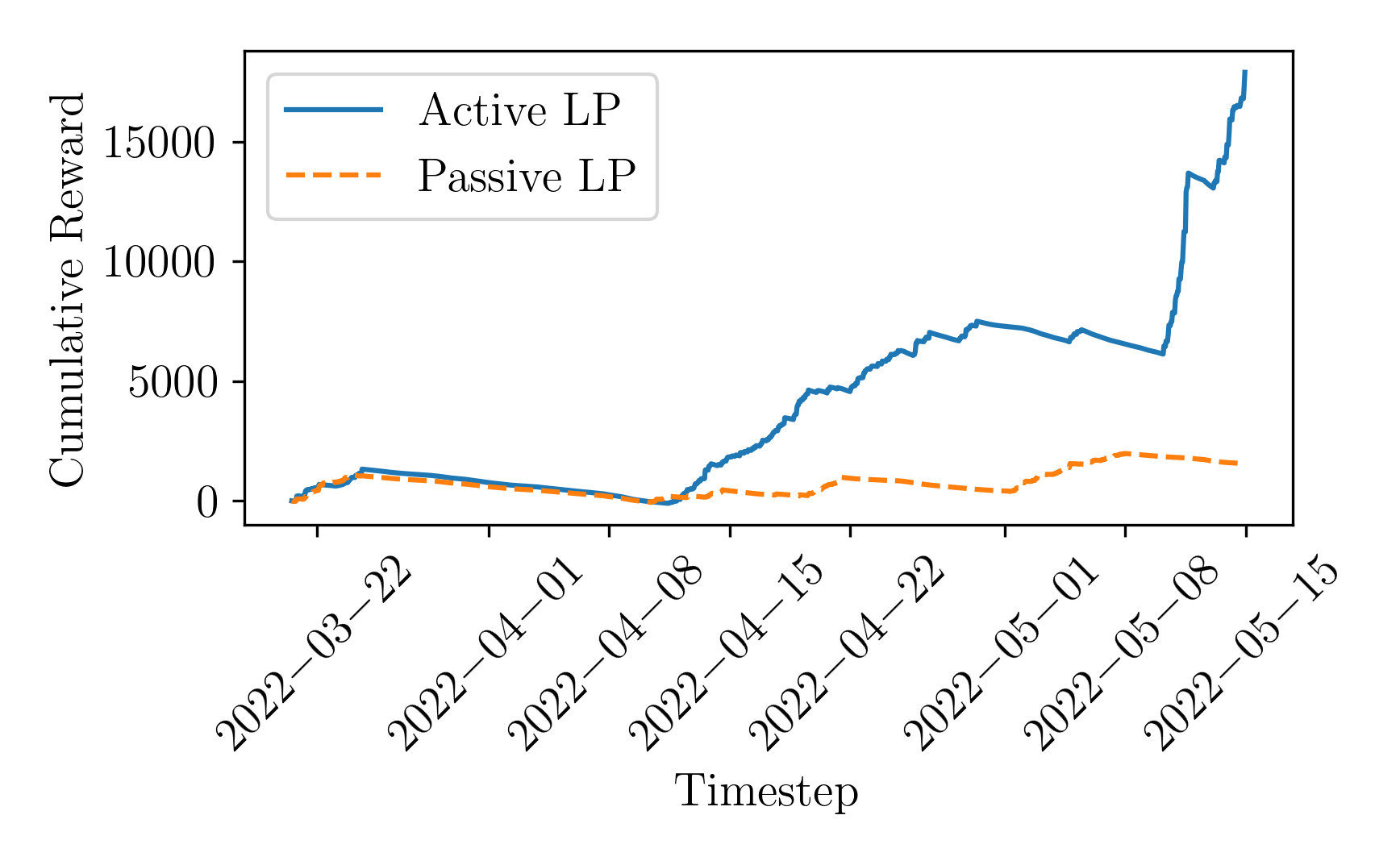}
    \end{subfigure}
    \hfill
    \begin{subfigure}[b]{0.48\linewidth}
        \centering
        \includegraphics[width=1.1\linewidth]{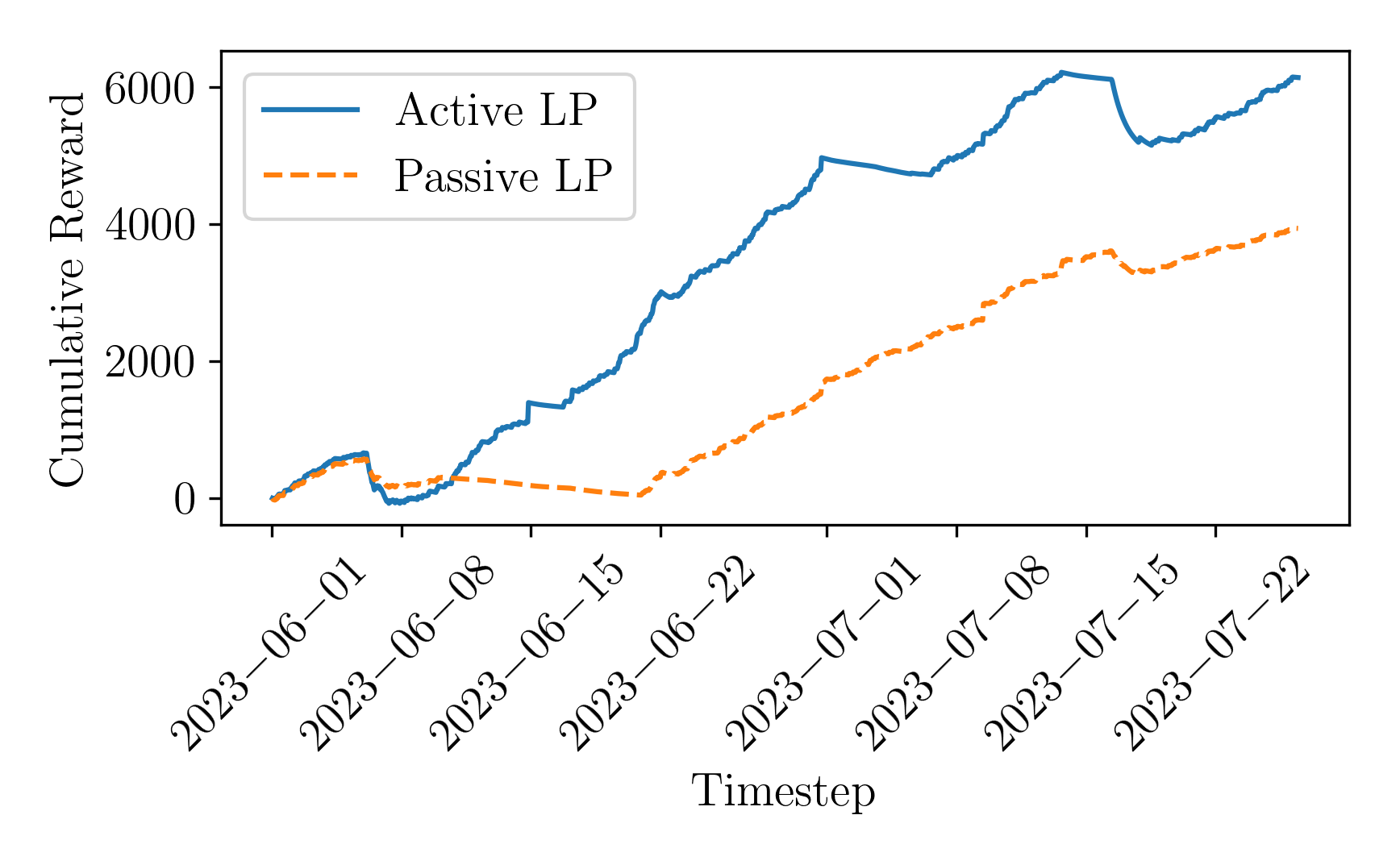}
    \end{subfigure}
    \caption{The two panels compare the cumulative rewards achieved by the active and passive LP strategies during the out-of-sample periods corresponding to the testing windows ending on 2022-05-14 (left panel) and 2023-07-26 (right panel). The plots illustrate the performance differences over time, emphasizing the active LP's ability to adapt to market dynamics and achieve higher rewards in varying conditions.}
    \label{Fig:cumrewards}
\end{figure}

Our results demonstrate that the active LP consistently outperforms the passive LP strategy in 7 out of the 11 out-of-sample windows. Fig. \ref{Fig:cumrewards} provides a detailed comparison with a table summarizing the cumulative rewards for both active and passive LP strategies across all windows, alongside a bar plot illustrating the differences. 

%%%%%%%%%%%%%%%%%%%%%%%%%%%%%%%%%%%%%%%%%%%%%%%%%%%%%%%%%%%%%%
\begin{figure}[h!]
    \centering
    % Adjust the table position
    \makebox[0.53\textwidth]{ % Adjust the width to shift the table
        \begin{tabular}{ccc}
        \toprule
            End of  Test & Active LP & Passive LP \\
            \midrule
            2022-05-14 & 17893.42 & 1555.46 \\
            2022-07-16 & 1546.03 & -1973.24 \\
            2022-09-16 & 834.13 & -493.90 \\
            2022-11-18 & 3160.38 & 2179.77 \\
            2023-01-19 & 3745.15 & 1149.32 \\
            2023-03-23 & 4953.87 & 4964.50 \\
            2023-05-24 & 3066.13 & 4939.25 \\
            2023-07-26 & 6143.00 & 3940.10 \\
            2023-09-26 & 4709.66 & 6047.50 \\
            2023-11-28 & 1356.69 & 4629.40 \\
            2024-01-29 & 6552.42 & 5780.83 \\
            \bottomrule
        \end{tabular}
    }
    % Plot below
    \includegraphics[width=0.5\textwidth]{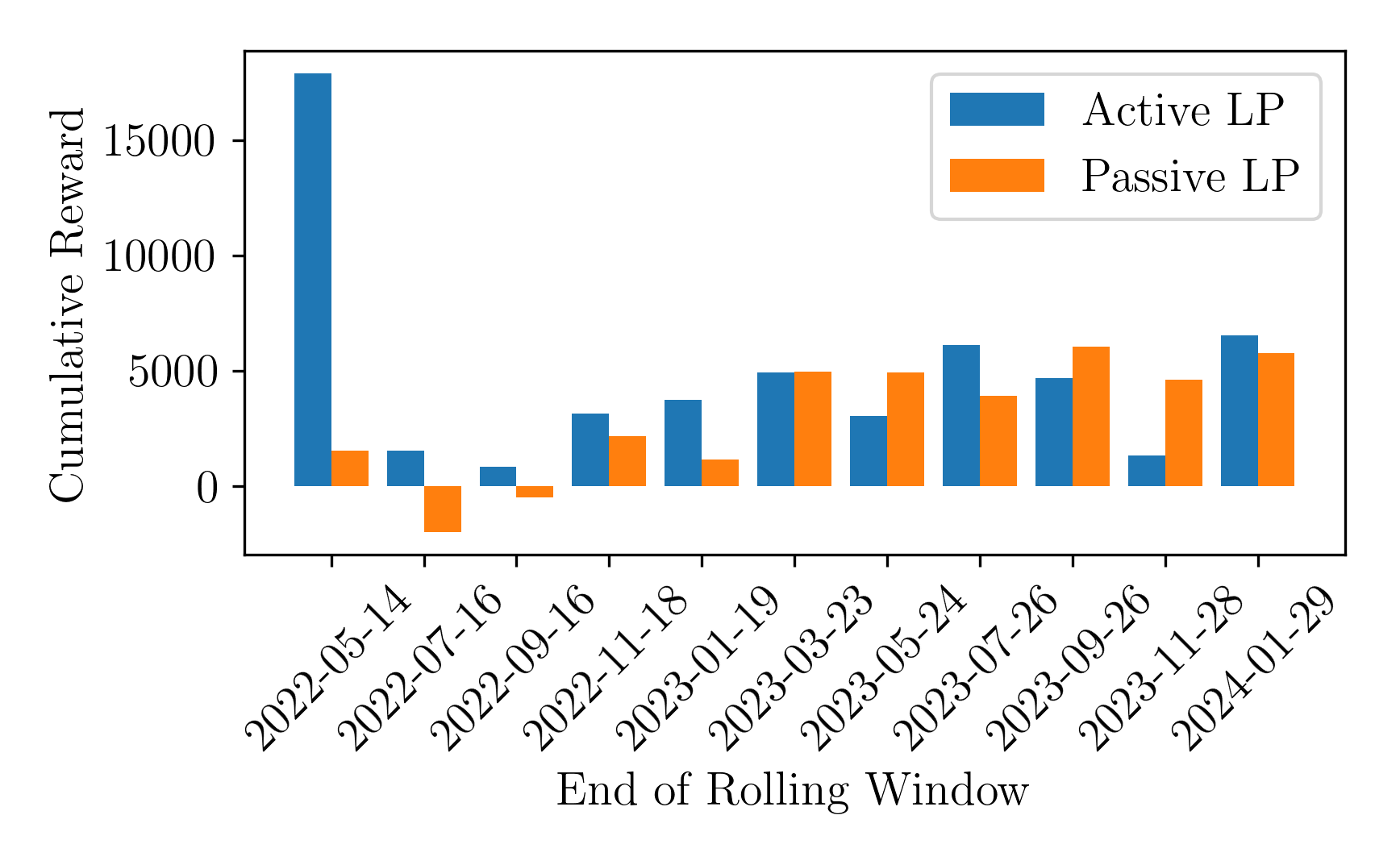}
    \caption{Comparison of active and passive LP strategies when $x_0=2$: The figure compares the cumulative rewards achieved by the active and passive LP strategies over the out-of-sample testing period. The bar plot in the bottom panel visually represents the performance differences, corresponding to the numerical values displayed in the table at the top.}
    \label{Fig:cumrew_table}
\end{figure}
%%%%%%%%%%%%%%%%%%%%%%%%%%%%%%%%%%%%%%%%%%%%%%%%%%%%%%%%%%%%%%

\begin{figure}[h]
    \centering
    \begin{subfigure}[b]{0.49\linewidth}
        \centering
        \includegraphics[width=1.1\linewidth]{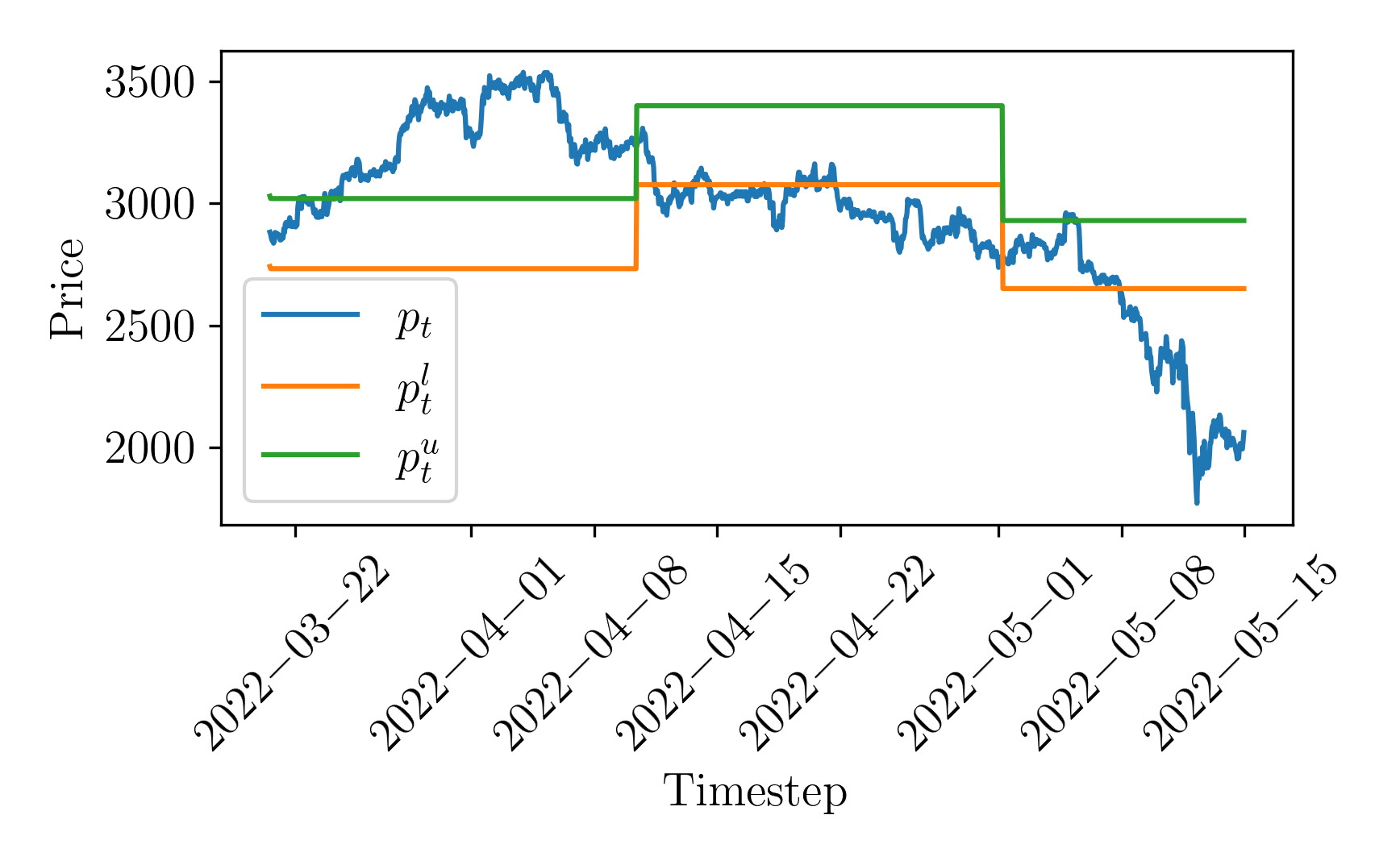}
    \end{subfigure}
    \hfill
    \begin{subfigure}[b]{0.49\linewidth}
        \centering
        \includegraphics[width=1.1\linewidth]{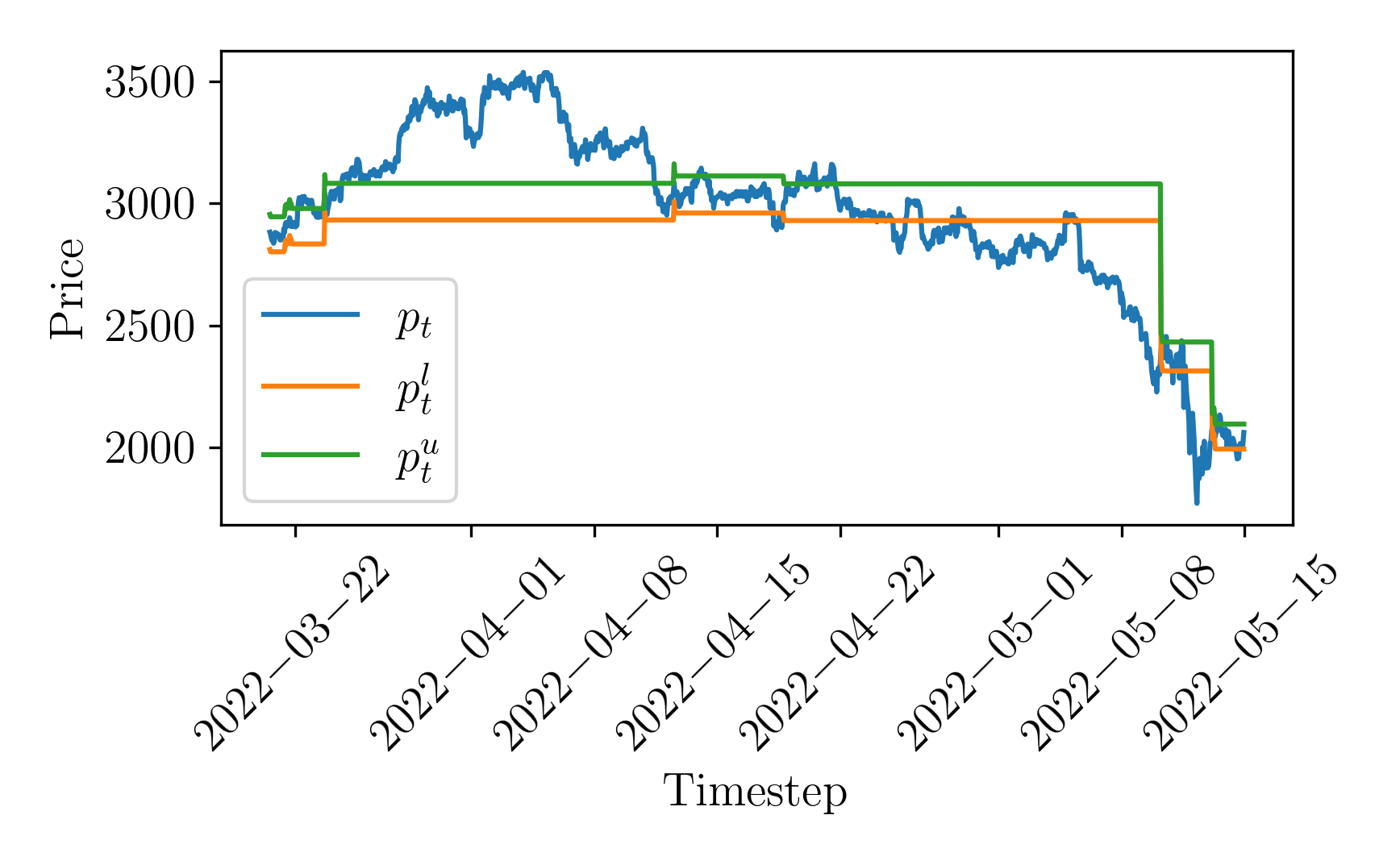}
    \end{subfigure}
    
    \vspace{0.5em} % Space between rows
    
    \begin{subfigure}[b]{0.48\linewidth}
        \centering
        \includegraphics[width=1.1\linewidth]{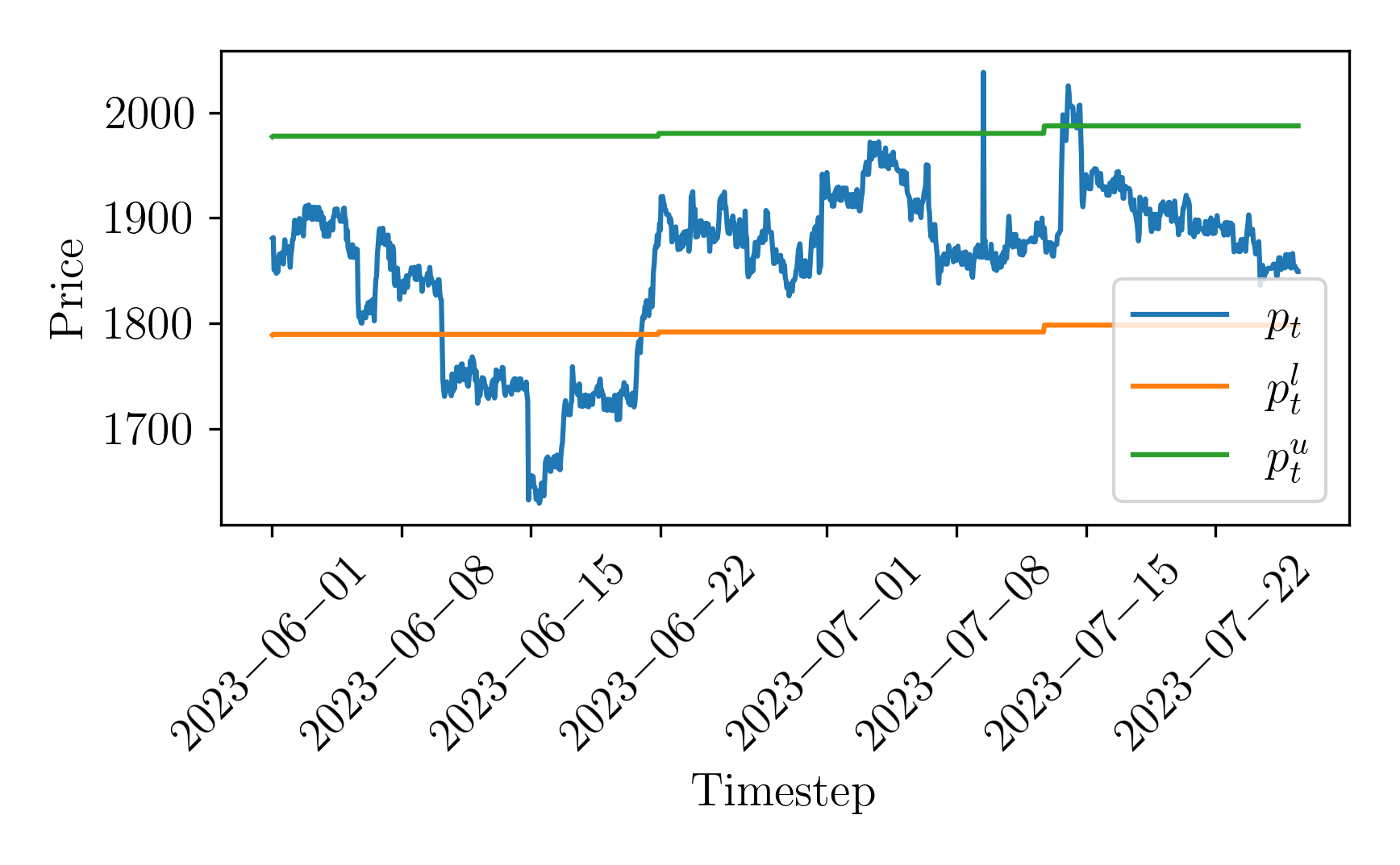}
    \end{subfigure}
    \hfill
    \begin{subfigure}[b]{0.48\linewidth}
        \centering
        \includegraphics[width=1.1\linewidth]{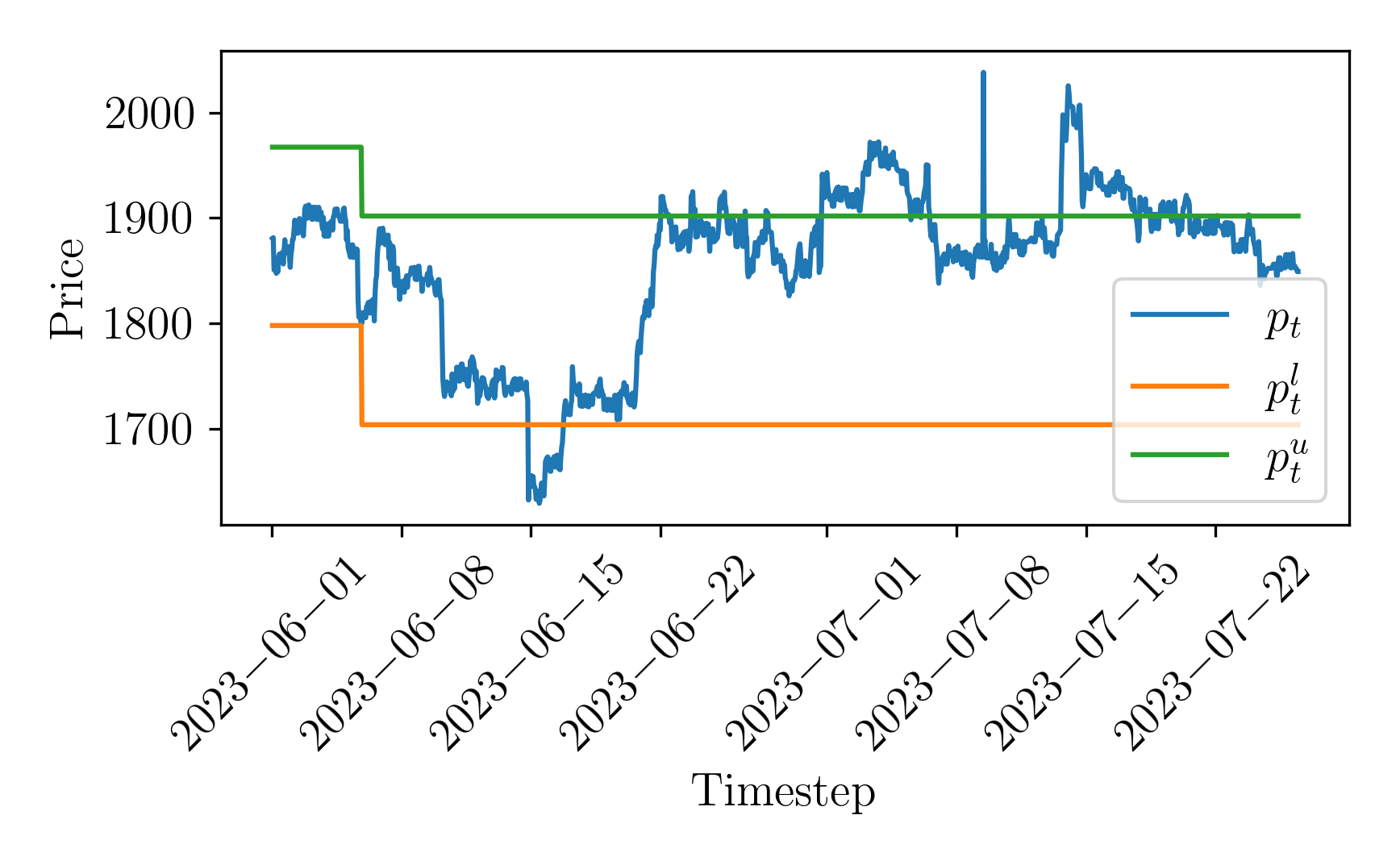}
    \end{subfigure}
    \caption{The figure presents four panels showing the WETH/USDC price dynamics in the Uniswap pool with a 0.05\% fee. The first column illustrates the price paths under a passive LP strategy, which rebalances at three predefined windows during the out-of-sample test period. The second column displays the equivalent price dynamics for the active LP strategy. The two windows correspond to the periods referenced in Fig. \ref{Fig:cumrew_table}, ending on 2022-05-14 (left column) and 2023-07-26 (right column).}
    \label{Fig:pricepaths}
\end{figure}

Fig. \ref{Fig:pricepaths} zooms in on two of the out-of-sample windows to compare the behavior of the passive LP and the active LP. As shown in the panels, the active LP learns to automatically adapt its price interval by leveraging the information provided in the state space. For instance, in the bottom-right panel, the active LP closely follows the price with narrower intervals but occasionally decides to keep the interval fixed, such as between the period of 2022-03-22 and 2022-04-08. During this time, the active LP anticipates a potential mean reversion in the trend, ensuring that the liquidity remains active. This illustrates the active LP's objective of optimizing the cumulative reward over time, avoiding unnecessary losses from frequent rebalancing and redeploying of positions.

In contrast, the top-left panel shows the behavior of the passive LP during the same time period. The price remains outside the liquidity range for most of the interval, leading to reduced performance. This is evident in Fig. \ref{Fig:cumrewards} (left panel), which displays the cumulative rewards over time, highlighting the superior performance of the active LP.

The second out-of-sample window in Fig. \ref{Fig:pricepaths} demonstrates a relatively stationary period for the token pair's price. In this case, the active LP (bottom-right panel) redeploys liquidity only once, recognizing the stationary nature of the price. Conversely, the passive LP (bottom-left panel) adheres to its heuristic rebalancing strategy, which results in unnecessary rebalancing and lower cumulative rewards, as observed in the cumulative reward panel on the right panel of Fig. \ref{Fig:cumrewards}.

%%%%%%%%%%%%%%%%%%%%%%%%%%%%%%%%%%%%%%%%%%%%%%%%%%%%%%%%%%%%%%
%%%%%%%%%%%%%%%%%%%%%%%%%%%%%%%%%%%%%%%%%%%%%%%%%%%%%%%%%%%%%%

%%%%%%%%%%%%%%%%%%%%%%%%%%%%%%%%%%%%%%%%%%%%%%%%%%%%%%%%%%%%%%
%%%%%%%%%%%%%%%%%%%%%%%%%%%%%%%%%%%%%%%%%%%%%%%%%%%%%%%%%%%%%%

\section{Takeaways and Conclusions}\label{sec:conclusion}

This study demonstrates the potential of DRL to optimize liquidity provisioning in DeFi environments such as Uniswap v3. The active LP agent balances frequent rebalancing to maximize fee collection with maintaining inactive positions during anticipated mean-reversion periods, reducing unnecessary costs. This capability arises from the long-term optimization framework of reinforcement learning.

Our empirical results reveal that the active LP outperforms heuristic-based passive LP strategies in diverse market conditions, achieving higher cumulative rewards in 7 out of 11 out-of-sample windows. The agent's ability to dynamically adjust liquidity intervals based on state-space information highlights its capability in addressing real-world challenges in liquidity management.

Future research could explore incorporating additional informational signals into the state space, such as off-chain trade volumes and price trends, which might indicate arbitrage risks impacting the LP position's value. Similarly, snapshots of the Ethereum mempool could provide valuable insights into impending on-chain orders, enabling the LP to proactively manage liquidity positions in anticipation of price movements on the AMM. These enhancements could further refine decision-making and improve the profitability of active liquidity provision strategies.

\bibliography{biblio}

\clearpage % Forces a new page before the appendix
\appendix

\section{Hyperparameter Optimization and Training Setup\footnote{Code will be made available upon acceptance in a GitHub repository.}}\label{App}

During the training of the PPO algorithm, we use an early stopping procedure to prevent overfitting by halting training when performance improvements stagnated beyond a set threshold. The training was conducted over a total of 100,000 timesteps, with updates performed after every third of the training window (corresponding to intervals of 25,000 timesteps).

Hyperparameter optimization was carried out by training 50 agents in each iteration and selecting the agent with the best performance. The grid of hyperparameters considered during the optimization process is provided in Tab. \ref{tab:hyp1} and \ref{tab:hyp2}.

\begin{table}[h]
    \centering
\begin{tabular}{cccc}
\toprule
End of Test & $\mathcal{A}$ & Activation & Hidden Layers \\
\midrule
2022-05-14 & [0, 20, 50] & sigmoid & [6, 4] \\
2022-07-16 & [0, 10, 20] & sigmoid & [4, 2] \\
2022-09-16 & [0, 40, 50, 60] & sigmoid & [4, 2] \\
2022-11-18 & [0, 50, 100] & relu & [6, 6, 6] \\
2023-01-19 & [0, 10, 20, 30] & relu & [6, 4] \\
2023-03-23 & [0, 40, 50, 60] & relu & [4, 2] \\
2023-05-24 & [0, 10, 20, 30] & tanh & [8, 2] \\
2023-07-26 & [0, 40, 50, 60] & sigmoid & [8, 2] \\
2023-09-26 & [0, 20, 50] & sigmoid & [10, 2] \\
2023-11-28 & [0, 50, 100] & sigmoid & [6, 2] \\
2024-01-29 & [0, 20, 50] & tanh & [8, 4] \\
\bottomrule
\end{tabular}
    \caption{Optimized hyperparameters over the training window for each testing window. The table includes the action space ($\mathcal{A}$), the activation function used in the neural networks, and a list indicating the number (lenght of the list) and size of hidden layers (numbers in the list).}
    \label{tab:hyp1}
\end{table}

\begin{table}[h]
    \centering
\begin{tabular}{cllll}
\toprule
 End of Test & LR & Clip & $c2$ & $\gamma$ \\
\midrule
2022-05-14 & 0.00005 & 0.05 & 0.00001 & 0.999 \\
2022-07-16 & 0.00005 & 0.2 & 0.0001 & 0.999\\
2022-09-16 & 0.00001 & 0.05 & 0.00001 & 0.999 \\
2022-11-18 & 0.01 & 0.1 & 0.0001 & 0.999 \\
2023-01-19 & 0.005 & 0.2 & 0.01 & 0.999 \\
2023-03-23 & 0.0001 & 0.05 & 0.0001 & 0.99 \\
2023-05-24 & 0.001 & 0.4 & 0.001 & 0.9 \\
2023-07-26 & 0.00001 & 0.4 & 0.00001 & 0.9 \\
2023-09-26 & 0.00001 & 0.05 & 0.01 & 0.9999 \\
2023-11-28 & 0.00001 & 0.1 & 0.00001 & 0.999 \\
2024-01-29 & 0.0001 & 0.2 & 0.0001 & 0.9999 \\
\bottomrule
\end{tabular}
    \caption{Optimized hyperparameters over the training window for each testing window. The table includes the learning rate (LR), the clip range of the PPO objective (Clip), the entropy coefficient in the objective ($c2$), and the discount factor ($\gamma$).}
    \label{tab:hyp2}
\end{table}

\section{Additional Results with Higher Liquidity}

\begin{figure}[h!]
    \centering
    % Adjust the table position
    \makebox[0.53\textwidth]{ % Adjust the width to shift the table
        \begin{tabular}{ccc}
        \toprule
            End of  Test & Active LP & Passive LP \\
            \midrule
            2022-05-14 & 17893.42 & 1555.46 \\
            2022-07-16 & 1546.03 & -1973.24 \\
            2022-09-16 & 834.13 & -493.90 \\
            2022-11-18 & 3160.38 & 2179.77 \\
            2023-01-19 & 3745.15 & 1149.32 \\
            2023-03-23 & 4953.87 & 4964.50 \\
            2023-05-24 & 3066.13 & 4939.25 \\
            2023-07-26 & 6143.00 & 3940.10 \\
            2023-09-26 & 4709.66 & 6047.50 \\
            2023-11-28 & 1356.69 & 4629.40 \\
            2024-01-29 & 6552.42 & 5780.83 \\
            2024-04-01 & -4405.51 & 1456.76 \\
            \bottomrule
        \end{tabular}
    }
    % Plot below
    \includegraphics[width=0.5\textwidth]{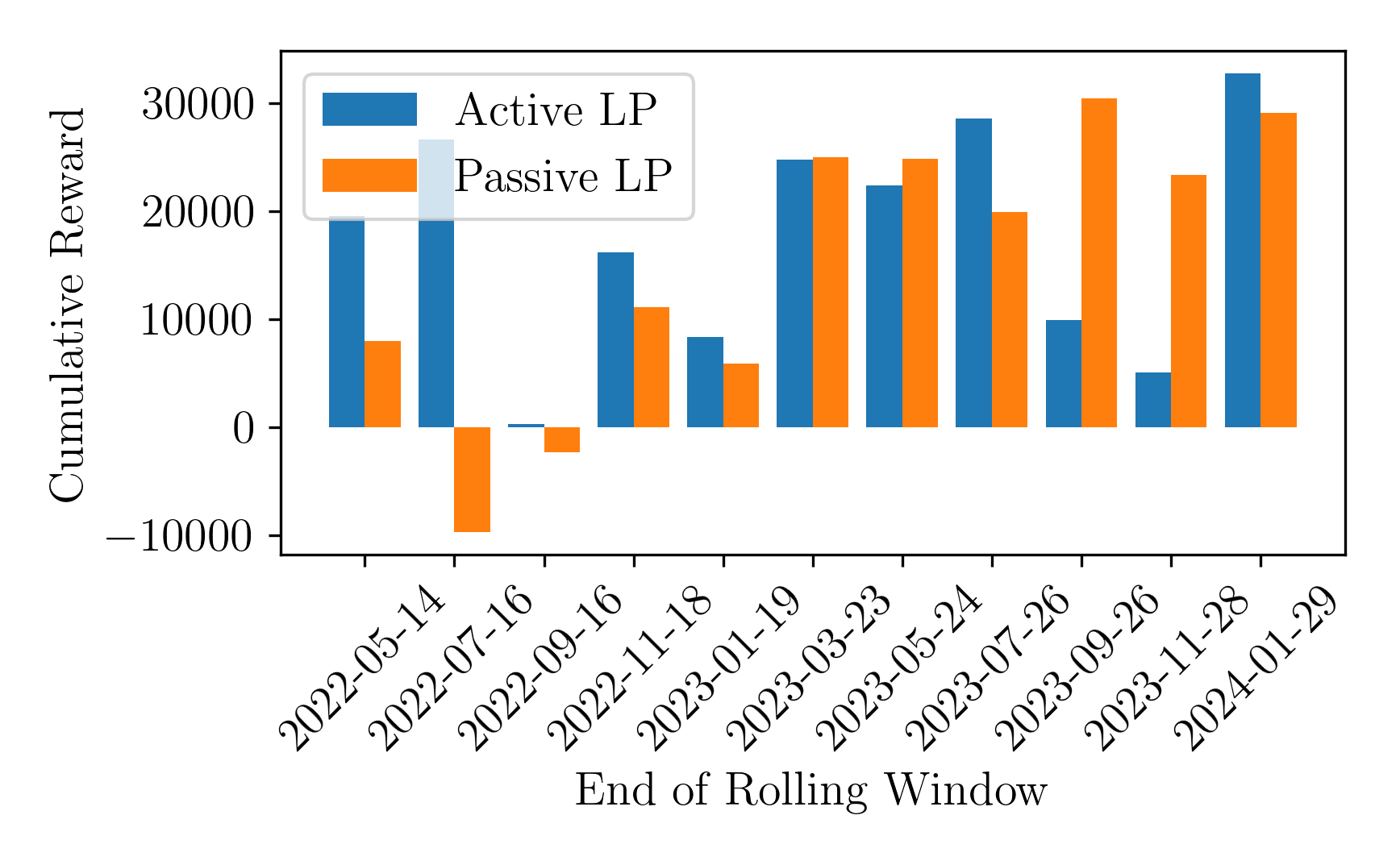}
    \caption{Comparison of active and passive LP strategies when $x_0=10$: The figure compares the cumulative rewards achieved by the active and passive LP strategies over the out-of-sample testing period. The bar plot in the bottom panel visually represents the performance differences, corresponding to the numerical values displayed in the table at the top.}
    \label{Fig:cumrew_tablex10}
\end{figure}

We evaluate the trained policies in scenarios where the initial liquidity increases from $x_0=2$ to $x_0=10$. This section provides results similar to those discussed earlier, illustrating the performance of both active and passive LP strategies with higher initial liquidity. The results align with those for $x_0=2$, as the active LP strategy outperforms the passive LP in 7 out of 11 testing windows. Fig. \ref{Fig:cumrew_tablex10} compares the cumulative rewards of the active and passive strategies during the out-of-sample testing period, emphasizing the influence of increased liquidity.

%%%%%%%%%%%%%%%%%%%%%%%%%%%%%%%%%%%%%%%%%%%%%%%%%%%%%%%%%%%%%%
%%%%%%%%%%%%%%%%%%%%%%%%%%%%%%%%%%%%%%%%%%%%%%%%%%%%%%%%%%%%%%

\end{document}